\documentclass[10pt,preprint2]{aastex}

\usepackage{graphicx}

\begin{document}

\title{The B and Be States of the Star EM Cepheus}

\author{Diana Kjurkchieva,\altaffilmark{1} Dragomir Marchev,\altaffilmark{1}
T.\ A.\ A.\ Sigut,\altaffilmark{2,3}
and Dinko Dimitrov\altaffilmark{4}}

\altaffiltext{1}{Department of Physics, Shumen University, 9700 Shumen, Bulgaria}
\altaffiltext{2}{Department of Physics \& Astronomy, The University of Western Ontario,
London, Ontario, Canada}
\altaffiltext{3}{Centre for Planetary Science and Space Exploration, The University of
Western Ontario, London, Ontario, Canada}
\altaffiltext{4}{Institute of Astronomy and NAO, Bulgarian Academy of Sciences,
Tsarigradsko shossee 72, 1784 Sofia, Bulgaria}

\begin{abstract}

We present eleven years of high-resolution, spectroscopic
observations for the star EM~Cep. EM~Cep
switches between B and Be star states, as revealed by the level of
H$\alpha$ emission, but spends most of its time in the B~star
state.  EM~Cep has been considered to be
an eclipsing, near contact binary of nearly equal-massed B stars
in order to reproduce regular photometric variations; however, this model
is problematic due to the lack of any observed Doppler shift in
the spectrum. Our observations confirm that there
are no apparent Doppler shifts in the wide spectral lines
H$\alpha$ and He\,{\sc i}~6678 in either the B or Be star states.
The profiles of He\,{\sc i}~6678 typically exhibited a filled-in
absorption core, but we detected weak emission in this line during
the highest Be state. Given the lack of observed Doppler shifts,
we model EM~Cep as an isolated Be star with a
variable circumstellar disk. We can reproduce the
observed H$\alpha$ emission profiles over the eleven year period
reasonably well with disk masses on the order of $3$ to
$10\times\,10^{-11}\;M_*$ in the Be state with the circumstellar
disk seen at an inclination of $78^{\circ}$ to the line of sight.
From a disk ejection episode in 2014, we estimate a mass
loss rate of $\approx\,3\times 10^{-9}\;\rm M_{\odot}\,yr^{-1}$.
The derived disk density parameters are typical of those found for
the classical Be stars. We therefore suggest that the EM~Cep
is a classical Be star and that its photometric variations
are the result of $\beta\;$Cep or non-radial pulsations. 

\end{abstract}

\keywords{stars: emission-line, Be: binaries: eclipsing; stars: individual (EM~Cep)\\
\vspace{0.02in}\\
Accepted for Publication in the Astronomical Journal, 10 June 2016}

\section{The Star EM Cepheus}

EM~Cep (HD~208392, HIP~108073, SAO~19718, GSC~04266-02575) is a
component of the visual binary ADS~15434 (Zasche et al.\ 2009) and a
member of the open cluster NGC~7160, with an age of 18 Myrs and
distance of 830 pc.

Emission in H$\alpha$ from EM~Cep was first reported by Plaskett
\& Pearce (1931) and Merrill et al.\ (1932). Rachkovskaia (1977)
did not find radial-velocity variations of the H$\alpha$ line
in excess of 15~km/s or any phase-dependent line-profile
variations. Hilditch et al.\ (1982) observed a H$\beta$ profile
with two emission peaks separated by 480~km/s, placed
symmetrically about a deep core. They found that the strengths of
H$\beta$ and H$\gamma$ sometimes changed by nearly a factor of
two within several hours but without a clear sign of duplicity
in the spectra. Uesugi \& Fukuda (1982) found the atmosphere of
EM~Cep to be helium-rich. EM~Cep is classified as a B1~III-IV
shell star with weak emission, and its $v\sin i $ is estimated to
be 280--290~km/s (Briot 1986, Hilditch et al.\ 1982).

EM~Cep was discovered to be a periodic variable by Lynds (1959),
with period of 0.8d and an amplitude of 0.15$^{m}$. Thus,
EM~Cep turned out a rare, short-period Be star (Pustylnik et al.\ 
2005). Due to its almost sinusoidal light curve, EM~Cep was
suspected to be a close-to-contact binary consisting of two
similar early-B stars. Further observations showed the light
variability to be quite irregular (Johnston 1970, Rachkovskaia
1975, 1976, Bakos $\&$ Tremko 1975). Flare events in the \emph{R}
band were also observed (Kochiashvili 1999, Kochiashvili et al.\ 
2007, Mars et al.\ 2010). Molik \& Wolf (2004) found that EM~Cep
falls on the ''blue envelope'' of the colour-period diagram of
close binaries.

The main peculiarity of EM~Cep is that its light variability has
not been found to be accompanied by any pronounced radial-velocity
or colour variations (Mars et al.\ 2010). Different explanations
have been proposed for this peculiarity: (a) an unstable common
envelope around the components of the contact binary (Johnston
1970, Bakos $\&$ Tremko 1975, Rachkovskaia 1977, Briot 1986); (b)
a magnetic, oblique rotator combined with a non-uniform brightness
distribution over the stellar surface (Harmanec 1984, Balona
1990); (c) pulsations of $\beta$~Cep or non-radial type
(Rachkovskaia 1976, Hilditch et al.\ 1982). Kochiashvili et al.\ 
(2007) suggested that the star reveals $\lambda\;$Eri-type
variability. Finally, we note that Pribulla $\&$ Rucinski~(2006)
found no observational confirmation of a third companion of
EM~Cep.

The search for a link between spectroscopic and photometric
variations is crucial for testing dynamic models of Be stars
(Hubert $\&$ Floquet, 1998). Our prolonged observations of EM~Cep
and their analysis aim to throw additional light on the nature of
EM~Cep and, perhaps, the Be phenomenon.

\section{The Classical Be Stars}

Classical Be stars are non-supergiant, early-type stars whose
spectra have, or have had at some time, one or more Balmer lines
in emission (Struve 1931). Decades of study have revealed that the
typically doubly-peaked Balmer emission lines originate from a
geometrically-thin, equatorial circumstellar disk fed by gas
ejected from the surface of rapidly-rotating B star (Porter $\&$
Rivinius 2003). There is also strong observational evidence that
these disks are rotationally supported with the disk matter in
Keplerian rotation around the central star (Meilland et al.\  2007,
Oudmaijer et al.\  2011, Meilland et al.\  2012, Wheelwright et al.\ 
2012).  The classical Be stars are particularly interesting due to
their fast rotation, with apparent equatorial speeds up to 70-80 $\%$ of
their critical velocity (Townsend et al.\  2004). It is possible
they have been spun up by mass transfer in binaries (McSwain $\&$
Gies 2005) or by the internal transport of angular momentum within
single stars (Granada et al.\ 2013).

Besides Balmer emission lines, Be stars are also characterized by
excess continuum emission at ultraviolet (UV), optical, and
infrared (IR) wavelengths (Gehrz et al.\  1974, Dachs et al.\  1988,
Kaiser 1989, Zorec $\&$ Briot 1991, Dougherty et al.\  1994,
Carciofi $\&$ Bjorkman 2006).

Presently, the family of classical Be stars numbers over 2000
known members as catalogued in the Be Star Spectra database
(BeSS1, Neiner et al.\  2011). Raddi et al.\ (2015) present a
catalogue of a further 247 photometrically and spectroscopically
confirmed fainter classical Be stars ($13^m \leq r \leq 16^m $) in
the direction of the Perseus Arm of the Milky Way.

The evolutionary status of classical Be stars is debatable
(Mermilliod 1982, Slettebak 1985). Martayan et al.\ (2007) found
that the evolution of the rotation speed with age is mass and
metallicity dependent and concluded that the Be phenomenon appears
earlier in the main sequence phase at higher stellar masses
($\sim$12 M$\odot$) and earlier spectral types, while it is
delayed in later B-types. On the other hand, observations of
Galactic-field, bright classical Be stars show a flat distribution
across the B sub-types (Zorec $\&$ Briot 1997).

The observational signatures of the Be stars, continuum excess and
line emission, are nowadays reproduced successfully by radiation
from a viscous decretion disk consisting of dense (n$_{e}\sim
10^{12}\,\rm cm^{-3}$) and hot (T$_e\sim 10\,000\,\rm K$) gas (Lee et
al.\ 1991, Porter 1999, Okazaki 2001, Bjorkman $\&$ Carciofi 2005,
Carciofi 2011, Carciofi et al.\ 2012).

\section{Spectral Observations of EM~Cep}

New spectral observations were carried out by the 2m RCC
telescope. We used a CCD Photometrics AT200 camera with an SITe
SI003AB 1024x1024 pixel chip mounted on the Coud\'{e} spectrograph
(grating B$\&$L632/14.7$^{\circ}$) on the 2m telescope of the
National Astronomical Observatory at Rozhen, Bulgaria. The
exposure time was 15~min. All stellar integrations were alternated
with Th-Ar comparison source exposures for wavelength calibration.
The resolution of the spectra is 16\,400, and most spectra have a S/N
ratio in the range 150--250. Initially, EM~Cep was observed in the
spectral range centered on H$\alpha$, but after July 2005, the
spectral range was changed to include the He\,{\sc i}~6678 line, 
$\rm 2p\,^1P^o - 3d\,^1D$.

The reduction of the spectra was performed using \emph{IRAF} (Tody
1993) packages for bias subtraction, flat fielding, cosmic ray
removal, and one-dimensional spectrum extraction. The spectra were
continuum normalized in the region around $6500\;$\AA.

Table~1 shows dates of our observations, the number of spectra,
and the observing phase calculated according to the ephemeris
\begin{equation}
JD({\rm Min\,I})=2452500.7420 + 0.806178\, E \;,
\end{equation}
from Kreiner et al.\ (2001).

\begin{table}
\caption{Journal of observations} \vspace{0.1in}
\begin{tabular}{lcl}
\hline\hline
Date         & Number of Spectra & Phase  \\
\hline
2004 Aug 24  & 18  &0.81-0.114 \\
2004 Aug 25  & 6   &0.04-0.43  \\
2005 Apr 15  & 11  &0.34-0.47  \\
2005 July 28 & 25  &0.11-0.47   \\
2005 July 29 & 1   &0.42        \\
2005 Aug 29  & 7   &0.76-0.87  \\
2005 Aug 30  & 4   &0.96-0.01  \\
2006 Feb 07  & 8   &0.74-0.83   \\
2006 June 21 & 2   &0.08-0.09  \\
2006 June 22 & 2   &0.27-0.29  \\
2006 July 18 & 10  &0.53-0.71  \\
2006 July 19 & 3   &0.76-0.79  \\
2007 Aug   8 & 5   &0.40-0.48  \\
2007 Dec 18  & 3   &0.05-0.18  \\
2007 Dec 19  & 2   &0.05-0.17  \\
2010 May 04  & 6   &0.7-0.77  \\
2010 May 07  & 4   &0.71-0.74 \\
2010 Sept 18 & 1   &0.78      \\
2011 Apr 11  & 1   &0.291     \\
2011 Sept 19 & 1   &0.60       \\
2011 Dec 04  & 1   &0.89    \\
2012 Jan 03  & 2   &0.96-0.99   \\
2012 Jan 04  & 1   &0.26        \\
2012 Jan 05  & 1   &0.42        \\
2012 June 30 & 1   &0.44      \\
2012 Aug 07  & 1   &0.41       \\
2012 Sept 08 & 2   &0.195      \\
2013 Apr 21  & 4   &0.35-0.39  \\
2013 Aug 25  & 3   &0.47-0.67   \\
2013 Oct 24  & 1   &0.83        \\
2013 Dec 18  & 2   &0.98-0.99   \\
2014 June 06 & 3   &0.096-0.121 \\
2014 July 8  & 2   & 0.64      \\
2014 July 9  & 1   & 0.81      \\
2014 Aug 08  & 2   & 0.12    \\
2014 Sept 30 & 2   &0.47-0.67   \\
2014 Nov 11  & 2   &0.47-0.67   \\
2014 Dec 13  & 1   &0.83       \\
2015 Jan 01  & 3   &0.98-0.99   \\
 \hline
\end{tabular}
\end{table}

\section{Qualitative analysis of the spectra}

Our 11 years of observations reveal that the
spectra of EM~Cep did not change significantly within a single
night but differed significantly over longer times. Figures A1 and A2 in the
Appendix chronologically show representative spectra from each observational
night, while Figures 1 and 2 show sample spectral corresponding
to the Be and B states.

\begin{figure}
\begin{center}
\includegraphics[width=8cm,scale=1.00]{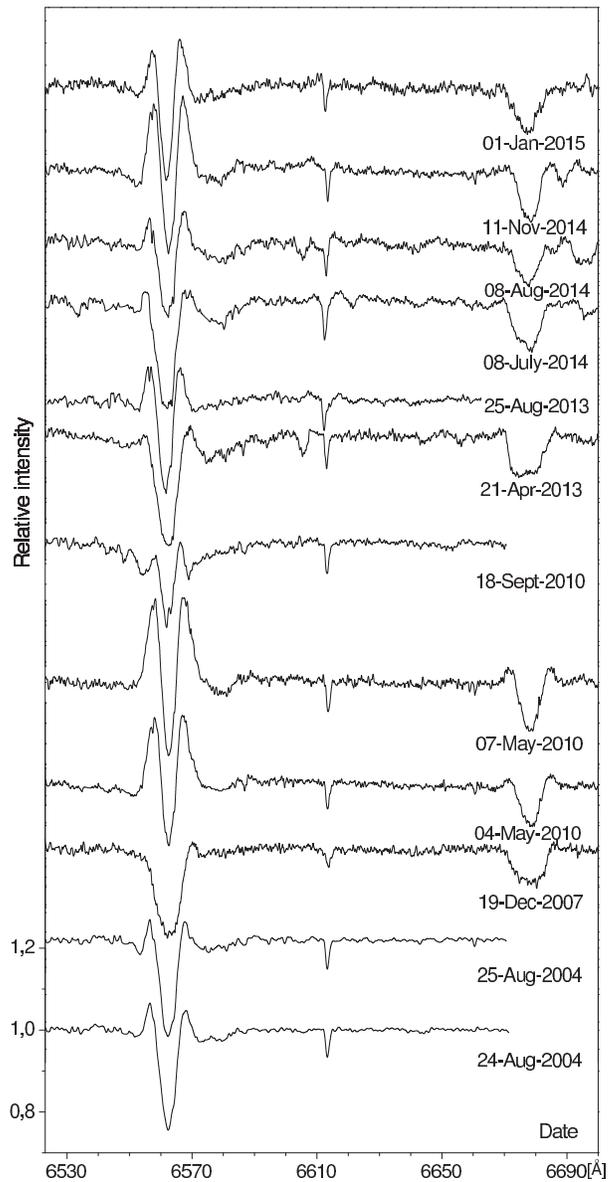}
\caption[]{A Sample of EM~Cep spectra in the Be state. The spectra
have been offset for clarity.}
\label{Fig1}
\end{center}
\end{figure}

\begin{figure}
\begin{center}
\includegraphics[width=7.8cm,scale=1.00]{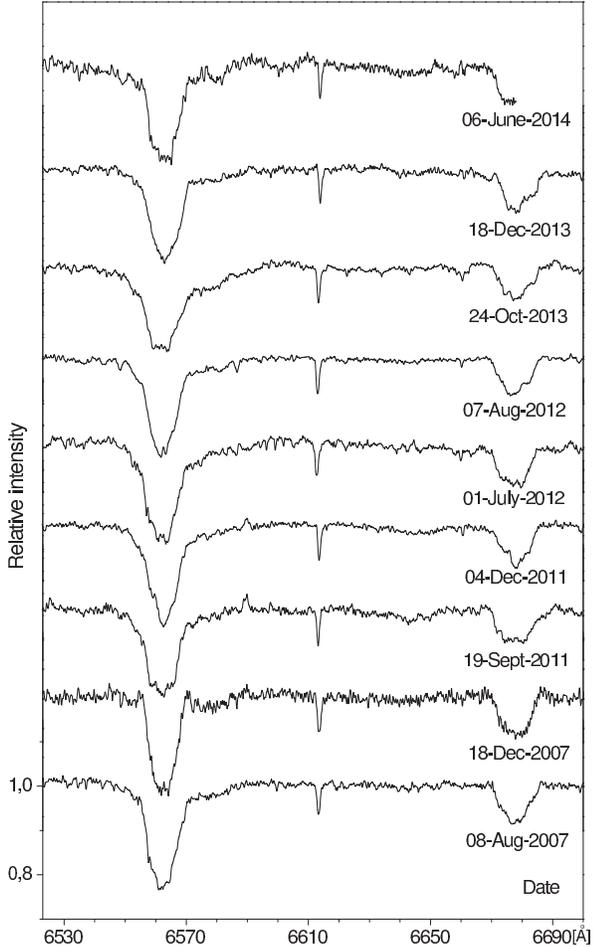}
\caption[]{A Sample of EM~Cep spectra in the B state.}
\label{Fig2}
\end{center}
\end{figure}

To quantify the behaviour of H$\alpha$ over the 11 year
period, we measured the following line parameters from our spectra:
(relative) intensities of the two emission
peaks in the Be state, I$_V$ and I$_R$; intensity of the line centre 
absorption core, I$_{abs}$; equivalent width (EW)  and
FWHM of the emission and absorption features; and the wavelengths
of the emission peaks in the Be state, $\lambda_V$ and $\lambda_R$.
All of these parameters for each observation date are given in Table~2.
In the B state, only the absorption component is present, and the
listed EW is the area measured below the continuum; the FWHM is
full width of the absorption line measured at half depth. In the Be states, an emission
EW is measured as the area in the wings above the continuum; the FWHM
of the emission component is the full width measured at half height of the emission wings,
whereas the FWHM of the absorption component refers to the full width at half depth of
the absorption core.

These line parameters allowed us to separate three levels of the Be state: 
(1) a high state with peak intensities above 1.1; (2) a middle state with peak
intensities between 1.05 and 1.1; and (3) a low state with peak
intensity below 1.05. In fact, the H$\alpha$ profiles at some low
Be states have only barely visible emission ``horns'' (or horn) above
the continuum or on the line wings (see 18 Sept.\ 2010 in Figure 1). All remaining
observations with no visible emission peaks were classified as the B state.
The variation of the classified state with time is shown in Figure~\ref{Fig3}, and
EM~Cep is seen to spend more time in the B~state than in the
Be~state.

We observed two high Be-state episodes: May 2010 and
November-December 2014. The spectra from May 4 and 7, 2010,
revealed that the H$\alpha$ emission increased considerably in the
framework of 3 days (see Figure~1 {and Table~2).

Our observations from April through August 2013, imply that the
Be state lasted at least 4 months, while observations from the
second half of 2014 allowed us to follow the transition from the B
state to the Be state and its end within 7 months (see Figure~1).

\begin{table*}
\caption{Measured line parameters for H$\alpha$.\label{tab:halpha}} \vspace{0.1in}
\begin{tabular}{lcccccccccc}
\hline\hline
Date & I$_V$    & I$_R$ &I$_{abs}$ & EW$_{em}$ & FWHM$_{em}$  & $\lambda_V-6500$ & $\lambda_R-6500$ & EW$_{abs}$ & FWHM$_{abs}$ & State\\
     &          &       &          & (\AA)    & (\AA)    & (\AA) & (\AA)        & (\AA) & (\AA) &        \\
\hline
2004 Aug 24     &  1.065& 1.047 &   0.755   & 0.304 &   14  &  56.49&  68.11&   1.153   &   5.8     &   Be middle\\
2004 Aug 25     &  1.05 & 1.044 &   0.764   & 0.148 &   13  &  56.47&  67.75&   1.211   &   5.7     &   Be middle\\
2005 Apr 15     &   -   &   -   &   0.811   &   -   &   -   &   -   &   -   &   2.279   &   12.1    &   B \\
2005 July 28    &   -   &   -   &   0.809   &   -   &   -   &   -   &   -   &   2.075   &   11      &   B \\
2005 July 29    &   -   &   -   &   0.778   &   -   &   -   &   -   &   -   &   2.213   &   9.7     &   B \\
2005 Aug 29     &   -   &   -   &   0.788   &   -   &   -   &   -   &   -   &   2.455   &   10.1    &   B \\
2005 Aug 30     &   -   &   -   &   0.784   &   -   &   -   &   -   &   -   &   2.015   &   9.8     &   B \\
2006 Feb 07     &   -   &   -   &   0.827   &   -   &   -   &   -   &   -   &   2.148   &   12.6    &   B \\
2006 June 21    &   -   &   -   &   0.808   &   -   &   -   &   -   &   -   &   1.99    &   11.1    &   B \\
2006 June 22    &   -   &   -   &   0.83    &   -   &   -   &   -   &   -   &   2.11    &   11.2    &   B \\
2006 July 18    &   -   &   -   &   0.795   &   -   &   -   &   -   &   -   &   2.082   &   10.1    &   B \\
2006 July 19    &   -   &   -   &   0.814   &   -   &   -   &   -   &   -   &   2.019   &   10.2    &   B \\
2007 Aug 8      &   -   &   -   &   0.764   &   -   &   -   &   -   &   -   &   2.477   &   10.5    &   B \\
2007 Dec 18     &   -   &   -   &   0.792   &   -   &   -   &   -   &   -   &   1.925   &   10      &   B \\
2007 Dec 19     &   -   & 1.014 &   0.792   & 0.065 &  -    &   -   & 70.435&  1.659    &   9.1     &   Be low\\
2010 May 04     &   1.16& 1.169 &   0.852   & 1.248 &   14  &  57.64& 67.199&  0.449    &   3.8     &   Be high\\
2010 May 07     &   1.2 & 1.206 &   0.82    & 1.691 & 15.2  &  57.638&67.539& 0.542     &   4.3     &   Be high\\
2010 Sept 18    &  0.961& 0.999 &   0.796   & 0.116 & 10.2  & 57.868& 66.238& 0.531     &   4.5     &   Be low \\
2011 Apr 11     &   -   &   -   &   0.82    &   -   &   -   &   -   &   -   &   2.374   &   12.5    &   B \\
2011 Sept 19    &   -   &   -   &   0.812   &   -   &   -   &   -   &   -   &   2.33    &   11.2    &   B \\
2011 Dec 04     &   -   &   -   &   0.775   &   -   &   -   &   -   &   -   &   2.452   &   10.4    &   B \\
2012 Jan 03     &   -   & 1.004 &   0.793   & 0.036 & -     &   -   & 70.884&  2.143    &   11.1    &   Be low \\
2012 Jan 04     &   -   & 0.983 &   0.777   & 0.02  & -     & -     & 71.622&  2.214    &   10      &   Be low \\
2012 Jan 05     &   -   & 1.013 &   0.783   & 0.033 & -     &  -    & 71.462&  2.14     &   10      &   Be low \\
2012 June 30    &   -   &   -   &   0.795   &   -   &   -   &   -   &   -   &   2.294   &   10.9    &   B \\
2012 Aug 07     &   -   &   -   &   0.781   &   -   &   -   &   -   &   -   &   2.351   &   10.6    &   B \\
2012 Sept 07    &   -   &   -   &   0.796   &   -   &   -   &   -   &   -   &   2.432   &   11.5    &   B \\
2013 Apr 21     & 1.005 & 1.023 &   0.734   &0.147  &   16  &  55.606& 70.48&  1.667    &   7.2     &   Be low\\
2013 Aug 25     & 1.083 & 1.079 &   0.773   &0.388  &   12.5& 55.788& 65.606& 0.879     &   4.6     &   Be middle \\
2013 Oct 24     &   -   &   -   &   0.82    &   -   &   -   &   -   &   -   &   2.615   &   11.5    &   B     \\
2013 Dec 18     &   -   &   -   &   0.803   &   -   &   -   &   -   &   -   &   2.093   &   11.4    &   B        \\
2014 June 06    &   -   &   -   &   0.793   &   -   &   -   &   -   &   -   &   2.169   &   11      &   B       \\
2014 July 8     & 1.024 & 1.028 &   0.741   & 0.107 & 15.6  & 55.052& 68.516& 1.67      &   6.7     &   Be low \\
2014 July 9     & 1.027 & 1.017 &   0.764   & 0.035 & 14.5  & 55.862& 68.308& 1.566     &    7      &   Be low\\
2014 Aug 08     & 1.064 & 1.081 &  0.825    &   0.4 & 13.6  & 56.436& 67.46 &   0.766   &   5.1     &   Be middle \\
2014 Sept 30    & 1.118 & 1.115 &  0.828    & 0.641 & 15.1  & 57.153& 67.093& 0.602     &   4.5     &   Be high \\
2014 Nov 11     & 1.163 & 1.182 &  0.795    & 1.213 &   14  & 57.638& 67.074& 0.631     &   3.75    &   Be high \\
2014 Dec 13     & 1.133 & 1.127 &  0.773    & 0.926 &  13   & 58.293& 67.093& 0.741     &   4       &   Be high \\
2015 Jan 01     & 1.082 & 1.108 &  0.791    & 0.505 & 12.5  & 57.257& 66.107& 0.711     &   4.15    &   Be middle \\
 \hline
\end{tabular}
\end{table*}

\begin{figure}
\begin{center}
\includegraphics[width=8.2cm,scale=1.00]{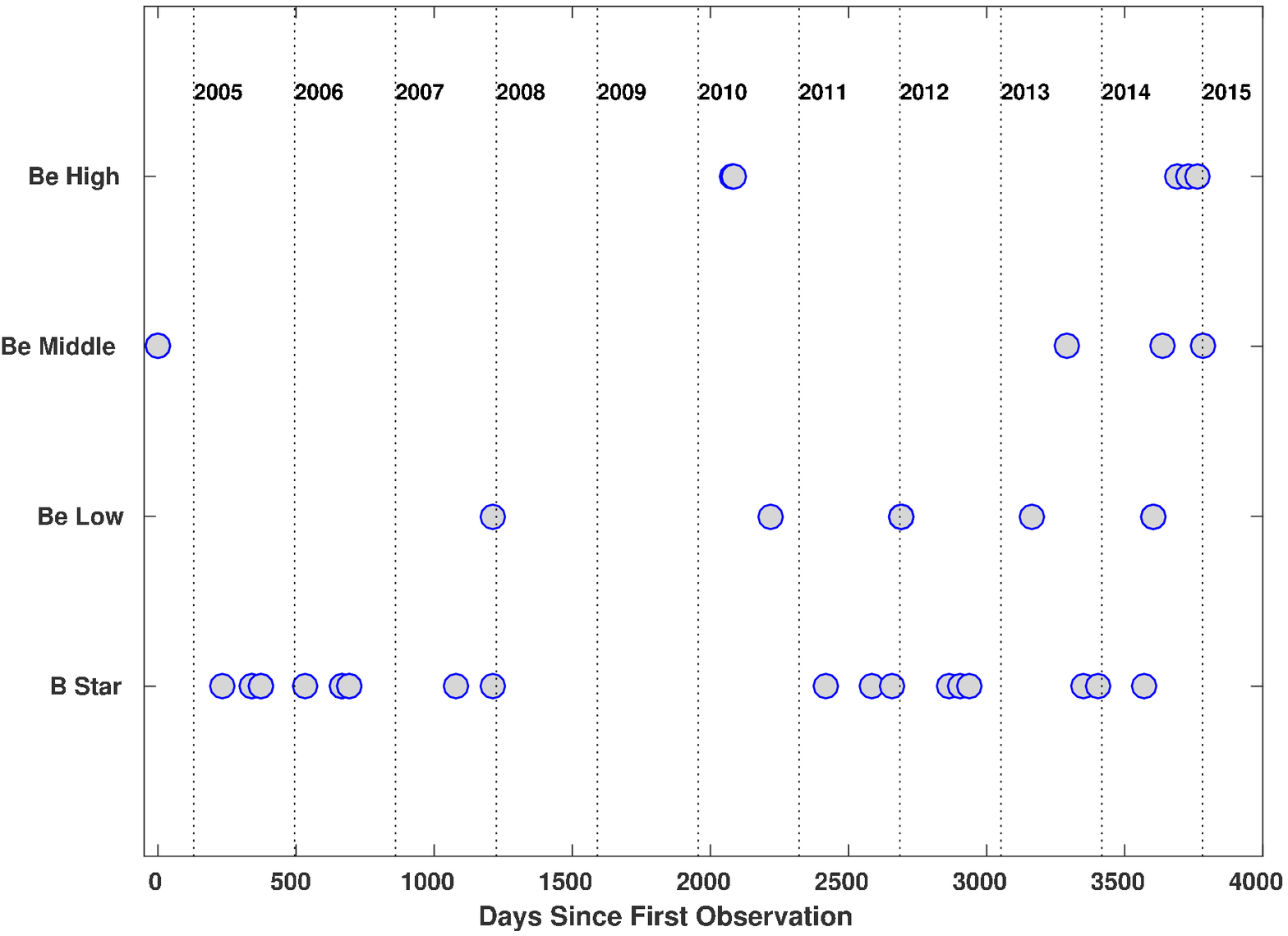}
\caption[]{State classification of EM~Cep (from Table~\ref{tab:halpha}) 
versus time. The year of each observing season is labelled at the top.}
\label{Fig3}
\end{center}
\end{figure}

\subsection{The H$\alpha$ line in the Be state}

The Be state H$\alpha$ profiles of EM~Cep consist of a deep, central
absorption core and two emission peaks (red and
blue-shifted peaks, R and V) that are almost symmetric with
respect to the absorption core (see Figure~1). This structure is
sunk in a shallow and very broad (nearly $40\;$\AA) absorption
trough. The absorption core and emission peaks are not precisely
in the middle of this broad absorption trough but closer to its
short-wavelength part. The bottoms of the H$\alpha$ absorption
cores are asymmetric as their right half is filled-in by emission
(Figure~1).

The mean separation between the R and V emission peaks is
$10\;$\AA\ at high Be state, $11\;$\AA\  at
middle Be state, and 12--$15\;$\AA\ at low Be state.  The
different separation of the two emission peaks of H$\alpha$
(Figure~1) means different velocities in the areas of the
strongest H$\alpha$ emission, i.e.\ the radii (and Keplerian
velocities) of the densest rings of the disk vary with time.

The depth of the absorption core tends to
deepen from high to low Be state, while its FWHM changes with the
H$\alpha$ emission strength, being $9-11\;$\AA\ at low Be state,
$4.5-7\;$\AA\  at middle Be state, and $3-5\;$\AA\ at high Be state.
However, the width of the whole H$\alpha$ profile (without the
very broad, shallow absorption trough) was always the same at
$\approx\,16\;$\AA.

The short-term variability during the 0.8d period of the
H$\alpha$ profiles in the Be state reveals two weak trends: the
depth of the absorption core is smaller at phases 0.0 and 0.5; and
the V peak is slightly higher than R peak in phase range 0.81-1.00
and vice versa in phase range 0.0-0.24. Harmanec~(1984) also found
the V/R ratio of H$\alpha$ to vary from 0.9 at phase 0.2 to 1.15
at phase 0.4.

The H$\alpha$ profiles in the Be states did not reveal any notable
trend of long-term, ``V/R'' variability (see Figure~1 and Table~2):
the two emission peaks had almost equal intensities in a
third of all Be states. The V peak was very rarely stronger than
the R peak, and the R peak was stronger than the V peak in about half of
the Be states, including low Be states when there was only
a small R peak, probably representing the first/last sign of the beginning/end of the
Be state. However, due to the small difference between the
intensities of the two emission peaks (up to 3\%, Table 2), EM Cep
cannot be considered a typical ``V/R variable'' Be star.

\subsection{H$\alpha$ in the B state}

The H$\alpha$ line of EM~Cep line in its B state is wholly in
absorption (Figure~2). However the profiles are slightly
asymmetric, with a filled-in core and equivalent width of
$1.9-2.6\;$\AA.

The short-term variability during the 0.8d cycle of the
H$\alpha$ profile in the B state is weak (Figure~4) while there is
a long-term variability of its depth and symmetry (Figure~2). The
spectra of Hilditch et al.\ (1982) also revealed variability of
the symmetry of the absorption Balmer lines H$\beta$ and
H$\gamma$.

We establish that the total width of the H$\alpha$ profile in the
Be and B states is almost the same (Figure~5). Their difference
represents the contribution of the additional H$\alpha$ emission
source at the Be state, a contribution that is obviously different
for the low, middle and high Be states. Its shape, illustrated in
Figure~5, clearly exhibits a disk-like origin.

The FWHM of the absorption H$\alpha$ line in B state is up
to 3.3 times larger than that of the H$\alpha$ absorption core at
Be state; however, their depths are the same (Table~2 and
Figure~5).

\begin{figure}
\begin{center}
\includegraphics[width=7.8cm,scale=1.00]{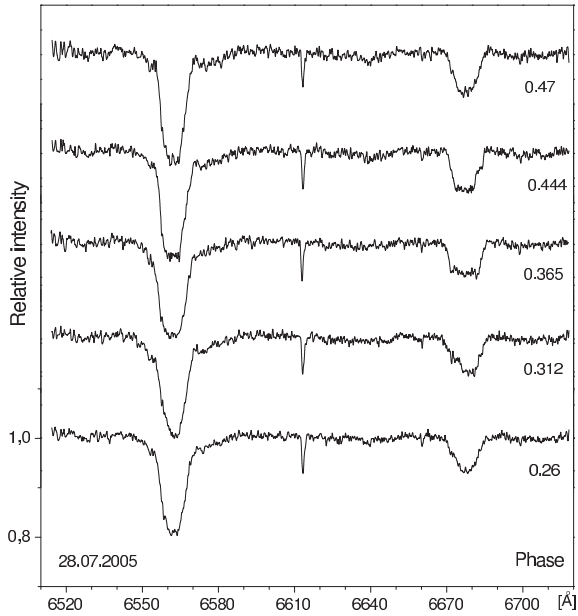}
\caption[]{Short-term variability of the spectra in the B state on
28 July 2005. The phases from Equation~1 are as indicated.}
\label{Fig2b}
\end{center}
\end{figure}

\begin{figure}
\begin{center}
\includegraphics[width=9.0cm,,scale=1.00]{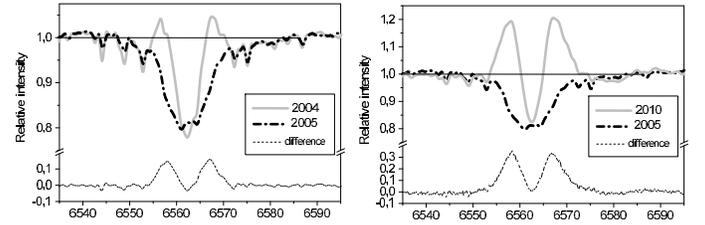}
\caption[]{H$\alpha$ in the B (dash-dot line) and Be states (solid
line). The difference between the Be and B state profiles are
shown below in each panel and represent the ``pure'' emission at
middle (left panel) and high (right panel) Be states.}
\label{Fig4}
\end{center}
\end{figure}

\subsection{The He\,{\sc i}~6678 line}

The depth of the He\,{\sc i}~6678 line was more than two times smaller
than that of H$\alpha$ (see Figures 1 and 2). Its
core is also filled-in, but in contrast to H$\alpha$, the He\,{\sc i}~6678
profile is not sunk in a broad trough. Translating its observed width,
$\approx\,\pm\,6.25\;$\AA\ from line centre, into a velocity shift gives
280~km/s, consistent with the $v\sin i $ estimate of Hilditch et al.\ (1982) 
obtained from the lines He\,{\sc i}~4388, He\,{\sc i}~4471, and Mg\,{\sc ii}~5581.

The short-term (phase) variability of the He\,{\sc i}~6678 profile
was more pronounced than that of H$\alpha$ (see Figure~4): it
becomes more symmetric and deeper at phases 0.0 and 0.5. Hilditch
et al.\ (1982) also noted that the He\,{\sc i}~4471 line changed more
regularly with the phase than the Balmer lines.

We found the following peculiarities in the behaviour of the
He\,{\sc i}~6678 line in 2005: it was wider and its right branch was
red-shifted by around 3~${\rm{\AA}}$ in the phase range
0.754-0.874 compared to the other phases, whereas the width and
position of H$\alpha$ remained almost the same; see Figure~6. We
noted the same, but weaker, effect in the phase range 0.14-0.23.
It is not clear if the whole He\,{\sc i}~6678 line is shifted to longer
wavelengths or only its right branch.

The strongest H$\alpha$ emission in May 2010 and November
2014 (high Be states) was accompanied with a weak emission in the
He\,{\sc i}~6678 line (see Figure~1). Then, its profiles
become more symmetric and its absorption core becomes deeper and
narrower. These observations represent the first detection of He\,{\sc i} emission for
EM~Cep. Table~3 gives the averaged measured line parameters for He\,{\sc i}~6678
at the B and Be states.

\begin{figure}
\begin{center}
\includegraphics[width=6cm,scale=1.00]{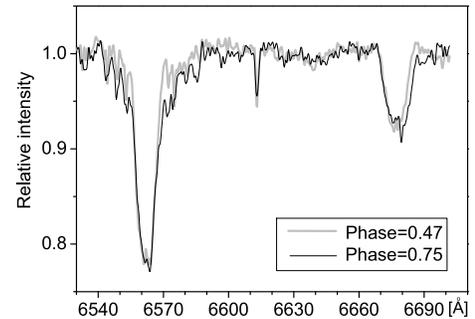}
\caption[]{The coincidence of the H$_{\alpha}$ profiles and the
difference of the position and the width of the He\,{\sc i} lines at
phases 0.757 and 0.47.} \label{Fig5}
\end{center}
\end{figure}

\begin{table*}
\caption{Averaged line parameters for He\,{\sc i} 6678 in the B and Be states.} \vspace{0.1in}
\begin{tabular}{lcccccc}
\hline\hline
State &  I$_{em}$ & I$_{abs}$ & EW$_{em}$ & EW$_{abs}$ & FWHM$_{em}$ & FWHM$_{abs}$\\
      &           &           & (\AA)      & (\AA)      & (\AA)      &   (\AA)       \\
 \hline
B   & -     & 0.914 & -    & 0.827 & -    & 11.7 \\
Be  & 1.036 & 0.882 & 0.19 & 0.676 & 16.2 & 7 \\
   \hline
\end{tabular}
\end{table*}

\subsection{Comparison of H$\alpha$ among different types of stars}

In Figure~7, we compare the H$\alpha$ profile of EM~Cep in its Be state
(Figure~7) with those of two other disk-like stars, FK~Com
(Kjurkchieva $\&$ Marchev 2005) and the nova-like, cataclysmic star
UX~UMa (Kjurkchieva et al.\ 2006), observed with the same
equipment. We find that all three H$\alpha$ profiles have very
similar widths and emission peak separations. Moreover, we found
that these profiles are similar also to the H$\beta$ and H$\gamma$
lines of the cataclysmic SU~UMa-type star HT~Cas (Catalan 1995),
as well as to the H$\alpha$ lines of some T Tau-type stars
(Muzerolle et al.\ 1998).

Therefore, we are faced the question as to why the width and emission
peak separation of the H$\alpha$ lines of different types of stars
are almost the same. Does this mean that there is some mechanism
of creating of similar disk structures with close velocities in
the different types of stars?

It should be noted that the Herbig Ae/Be stars are
also intermediate mass stars with circumstellar disks (Reipurth et al.\ 1996;
Cauley $\&$ Johns-Krull 2015). However, most of them exhibit
considerably stronger H$\alpha$ emission than EM Cep and their
profiles are strongly asymmetric, with considerable differences between
the V and R intensities). We have found only two Herbig Ae/Be stars
with H$\alpha$ profiles similar to those of EM Cep: AK Sco
(see Figure~1 in Reipurth et al.\ 1996) and CQ Tau (see Figure~19 in Cauley
$\&$ Johns-Krull 2015).

\begin{figure}
\begin{center}
\includegraphics[width=9cm,scale=1.00]{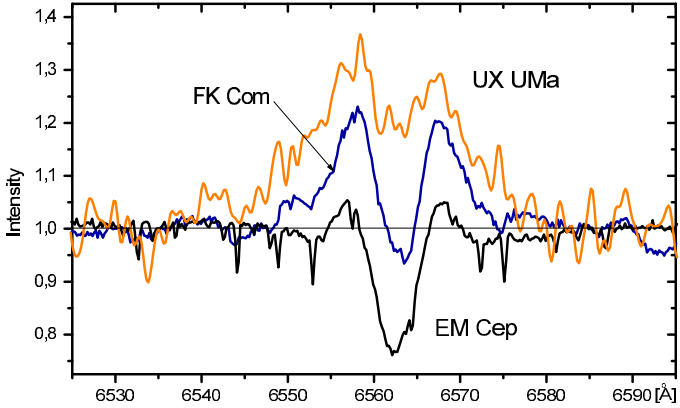}
\caption[]{The similarity of the H$\alpha$ lines of UX~UMa, FK~Com
and EM~Cep at Be state, observed with the same equipment}
\label{Fig11}
\end{center}
\end{figure}

%
%

\section{Modelling of the H$\alpha$ profiles of EM~Cep}

The 2004--2015 H$\alpha$ spectra of EM~Cep seem very similar to
typical Be stars in terms of strength and morphology of the
emission profile and in the long-term time variation of its
strength (for examples of Be star spectra, see Dachs~1988). The
H$\alpha$ profile of EM~Cep in its Be state shows approximate
symmetry between the red and blue emission in the wings over the
duration of the observations. In the maximum Be state observed for
EM~Cep (7~May~2010), the H$\alpha$ shell parameter, defined as the
average flux in the emission peaks divided by the flux at line
centre, is $\sim 1.5$, just satisfying the definition of a Be
shell star (Hanuschik 1996). Hence, the target is likely viewed at
a high inclination ($i>70^o$) relative to the common rotation axis
of the star and circumstellar disk so that the observer's line of
sight to the star first passes through the disk.

For this section, we will assume that EM~Cep is rapidly rotating,
main sequence, B star surrounded by a thin circumstellar disk in
Keplerian rotation (Porter $\&$ Rivinius 2003). Such a model has
been widely and very successfully used to model the emission
lines, infrared excesses, polarization, and interferometric images
of many Be stars (Rivinius et al.\ 2013a).

We modelled the H$\alpha$ line of EM~Cep as originating from such
a circumstellar disk surrounding a B1V star whose parameters are
given in Table~2. The equatorial surface density of the disk was
assumed to decrease as a power-law with the radial distance from
the star's rotation axis, $R$, as
\begin{equation}
\sigma(R) = \sigma_0 \left(\frac{R_{\rm eq}}{R}\right)^m \,
\label{eq:surfden}
\end{equation}
where $R_{\rm eq}$ is the star's equatorial radius. We have
considered 12 base disk surface densities, $\sigma_0=0.001$,
0.003, 0.006, 0.010, 0.033, 0.066, 0.100, 0.333, 0.666, 1.000,
3.333, 6.666, and $10.000\;\rm g\,cm^{-2}$, and five power-law
indexes, $m=1.0$, 1.5, 2.0, 2.5, and 3.0. Thus, a total of 60 disk
density models were considered.

%
%

\begin{table}
\begin{center}
\parbox{2.9in}{
\caption{Adopted stellar parameters for EM~Cep. \label{tab:param}}
}
\begin{tabular}{@{}ll}
\hline \hline
Parameter      \hspace{4cm}    &   Value \\
\hline
Mass$^{\rm a}$ ($M_{\odot}$)   &   12.5        \\
Polar Radius$^{\rm a}$ ($R_{\odot}$)      &  6.3  \\
Equatorial Radius$^{\rm a}$ ($R_{\odot}$)      &  9.45 \\
Luminosity ($L_{\odot}$)  & $1.6\times 10^4$ \\
$v_{\rm frac}$ & 0.95 \\
$\rm T_{eff}$(K)$^b$         & $26,\!000$        \\
$\log(g)^b$                & $4.0$            \\
\hline
\end{tabular}
\medskip\\
\parbox{2.9in}{
\footnotesize
{\bf Notes.}\\
$^{\rm a}$ Adopted from Townsend et al.\ (2004).\\
$^{\rm b}$ As defined by the luminosity and polar
radius.\\
}
\end{center}
\end{table}

\begin{figure}
\begin{center}
\includegraphics[scale=0.56]{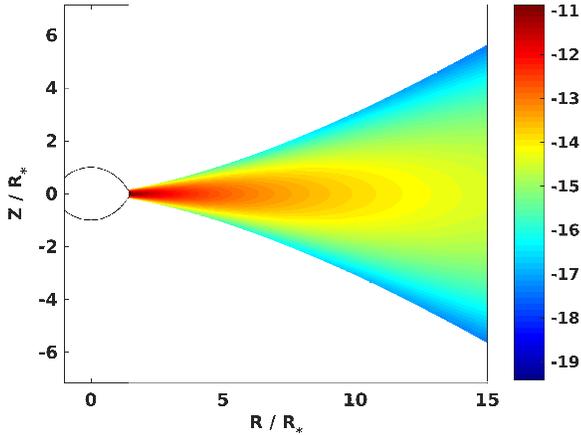}
\caption[]{The {\sc bedisk} disk density model corresponding
$\sigma_0=0.666\,\rm g\,cm^{-2}$ and $m=2.0$. The colours represent
$\log\rho(R,Z)$ according to the colour bar on the right, where the density 
is in units of $\rm g\,cm^{-3}$. The outline of the
central, distorted star is shown on the left.}
\label{fig:disk_temp}
\end{center}
\end{figure}

\begin{figure}
\begin{center}
\includegraphics[scale=0.46]{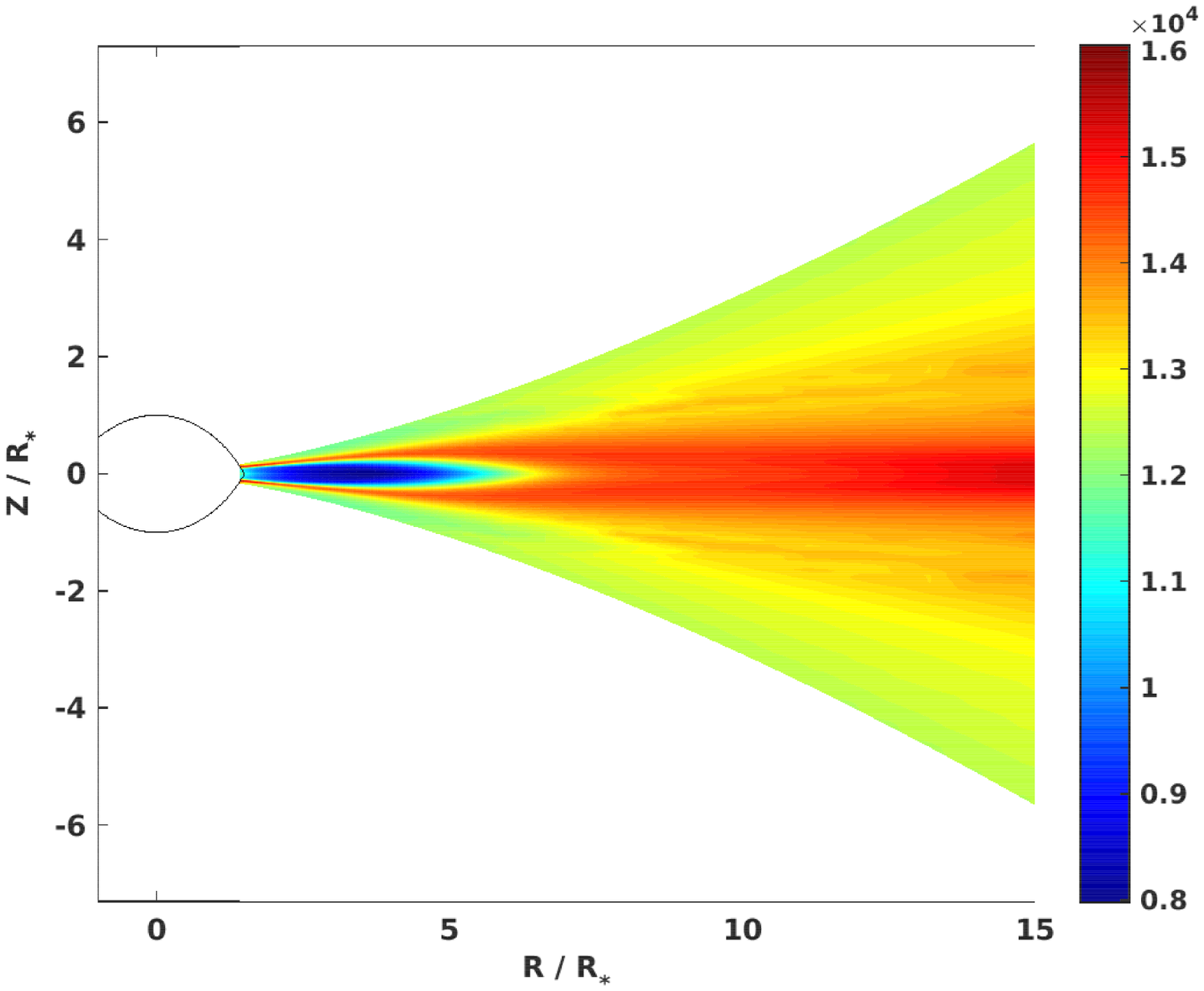}
\caption[]{The calculated {\sc bedisk} temperatures corresponding
$\sigma_0=0.666\,\rm g\,cm^{-2}$ and $m=2.0$. The colours represent
$T(R,Z)$ in degrees Kelvin according to the colour bar on the right.
The outline of the central, distorted star is shown, but not the 
temperature variation across its surface.}
\label{fig:disk_temp}
\end{center}
\end{figure}

We have used the code {\sc bedisk} (Sigut $\&$ Jones 2007) to
compute the temperature structure of the disk for each of the 60
density models. {\sc bedisk} assumes that the sole energy input
into the disk is the central star's photoionizing radiation field,
and it enforces radiative equilibrium by balancing the processes
of heating and cooling in a gas of solar chemical composition
consisting of the 9 most abundance elements (and their ions). {\sc
bedisk} produces both the temperature distribution in the disk and
atomic level populations that can be used later to formally solve
the radiative transfer equation to predict spectral lines of
interest, such as H$\alpha$. {\sc bedisk} assumes an axisymmetric
disk and returns the disk temperature $T(R,Z)$, where $Z$ is the
height above or below the equatorial plane.

To transform the surface density (Eq. (2)) into the corresponding
mass density $\rho(R,Z)$, it was assumed that at each radial
distance $R$, the disk is in vertical hydrostatic equilibrium with
a scale height $H$ parametrized by a single temperature $T_{0}$ as
\begin{equation}
H=\left(\frac{2R_*^3\;kT_{0}}{GM_*\;\mu_m
m_{\rm H}}\right)^{1/2} \, \left(\frac{R}{R_*}\right)^{3/2}\equiv
\beta(T_0)\,R^{3/2}\;. \label{eq:scale_height}
%
%
\end{equation}
Here $M_*$ and $R_*$ are the mass and radius of the central star,
and $\mu_m$ is the mean-molecular weight of the gas in the disk.
The hydrostatic temperature for EM~Cep's disk was set to
$T_0=0.6\,T_{\rm eff}$ (Sigut et al.\  2009), or 15600 K. This
choice gives $H/R_*=0.037$ at the inner edge of the disk. With
this scale height, the mass density in the disk is given by
\begin{equation}
\rho(R,Z) = \rho_o \left(\frac{R_*}{R}\right)^{n} \,
e^{-\left(\frac{Z}{H}\right)^2} \;,  \label{eq:rho}
\end{equation}
with
\begin{equation}
\rho_0=\frac{\sigma_0}{\sqrt{\pi}\beta(T_0) R_*^{3/2}} \,,
\end{equation}
and
\begin{equation}
n=m+\frac{3}{2}\;.
\end{equation}

As rapid rotation of the central star is though to be a key driver
of the Be phenomena (Rivinius 2013b), a gravitationally darkened
and a distorted central star should be considered (McGill et al.\ 
2011). Ejection of the circumstellar disk may be linked to
episodes of near critical rotation that occur naturally in hydrodynamic
models of rapidly rotating B type stars (Granada et al.\ 2013).
For EM~Cep, we assumed an equatorial rotation velocity of
95\% of the critical velocity. The distortion of the stellar
surface due to the rapid rotation was computed in the Roche
approximation (Collins 1966) and the temperature variation from
the pole to equator (i.e.\ the gravitational ``darkening") was
computed with the Espinosa Lara $\&$ Rieutord (2011) formalism.

The assumption of near critical rotation for EM~Cep, coupled with the
likely large viewing inclination of the disk (noted above and found
from the detailed modelling below), seems to lead to an inconsistency with
EM~Cep's $v\sin i$ of $\approx 280\,\rm km\,s^{-1}$. However,
strong gravitational darkening is known to lead to a systematic underestimation of
$v\sin i$ based on spectral lines because the most rapidly rotating portions
of the stellar disk are darker (see Townsend et al.\ 2004).
Finally, we note that the stellar $v\sin i$ has only a small effect on the
shape of the strongly Stark-broadened photospheric H$\alpha$ line.

\begin{figure}
\plotone{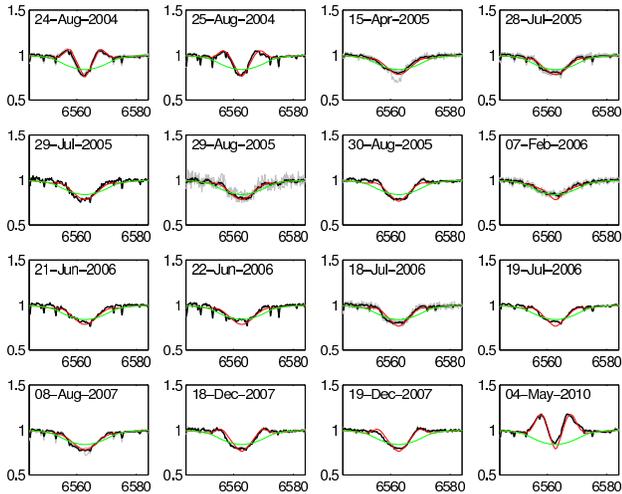}
\caption{The observed H$\alpha$ profiles (black), the best-fitting
{\sc beray} model profiles, and photospheric H$\alpha$ profiles
for the central star (green) for spectra between 24~August~2004
and 4~May~2010. \label{fig:stacksub:a}}
\end{figure}

\begin{figure}
\plotone{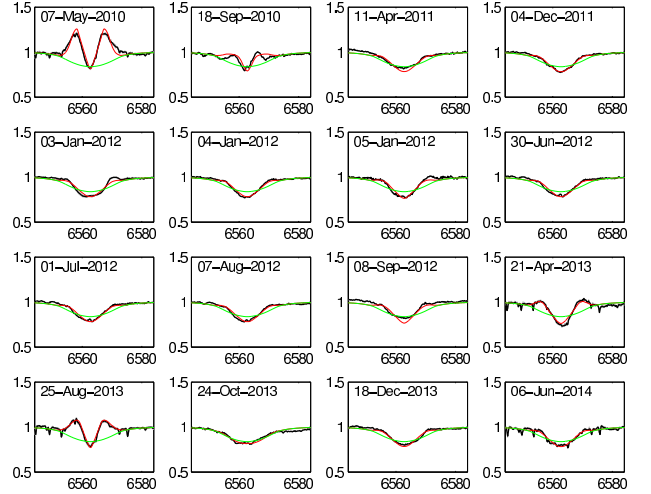}
\caption{Same as Figure~8 but for spectra between 7~May~2010
and 6~June~2014.  \label{fig:stacksub:b}}
\end{figure}

\begin{figure}
\plotone{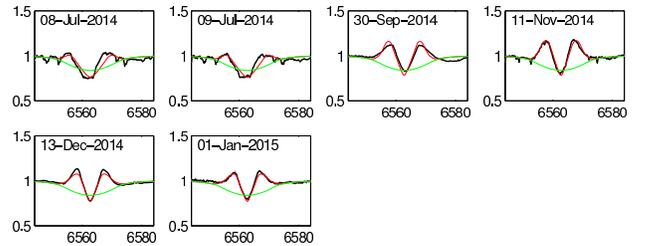}
\caption{Same as Figure~8 but for spectra between 6~July~2014
and 1~January~2015. \label{fig:stacksub:c}}
\end{figure}

The density structure of a disk with parameters 
$\sigma_{0}=0.666\,g\,cm^{-2}$ and $m=2$ is shown in Figure~8, and the 
corresponding disk temperature distribution obtained by {\sc bedisk} 
is shown in Figure~9. The circumstellar disk is far from isothermal, with a
cool, inner region close to the star resulting from the large
optical depths along all rays back to the central star. The strong
temperature variation across the surface of the star, from $T_{\rm
eff}$=27200 K at the pole to $T_{\rm eff}$=18300 K at the equator,
is not shown in Figure~9 in order to highlight the (lower)
temperatures in the disk. Discussion of the temperature structure
of Be star disks can be found in Sigut $\&$ Jones (2007), McGill
et al.\  (2011, 2012).

To compute predicted H$\alpha$ line profile corresponding to each
of the disk density models, the {\sc beray} code of Sigut~(2011)
was used. {\sc beray} formally solves the equation of radiative
transfer along rays threading the star-disk system directed at the
observer. Rays that terminate on the stellar surface use an LTE,
photospheric, H$\alpha$ profile, appropriate to the local $T_{\rm
eff}$ and $\log\,g$ on the stellar surface as the boundary
condition, shifted by the radial velocity of that patch. Rays that
pass entirely through the disk use a zero boundary condition. The
{\sc beray} modelling adds two additional parameters: the viewing
inclination of the system (values of $i=18$, $31$, $45$, $60$,
$72$, $75$, $78$, $81$, $84$, $87$ and $90^o$ were used) and the
outer disk radius (values of $R_d=6$, $12$, $25$ and $50\,R_*$
were considered). In all cases, the circumstellar disk was assumed
to start at the stellar surface, i.e.\ $R = R_{\rm eq}$, and
extend to $R_d$.

In total, a library of 2640 H$\alpha$ profiles was computed for
comparison with the observations. All profiles in the library were
convolved down to a resolution to match the observations.
To fit the appropriate model to each observation, a median
observed H$\alpha$ profile was created for each observing date,
and this profile was then compared to the H$\alpha$ library. The
best-fit profile was chosen by minimizing
\begin{equation}
{\cal F} \equiv \frac{1}{N}\sum_i \frac{|F_i^{\rm mod}
                - F_i^{\rm obs}|}{F_i^{\rm obs}} \;,
\label{eq:fom}
\end{equation}
where the sum over $i$ is for all $N$ wavelengths in the range
$6550\le\lambda_i\le 6570$.

In total, 38 such best-fit profiles were generated for the 38
individual days of data, spanning observations from 2004 through
2015. Initially the viewing inclination was left as a free
parameter, and the results showed that a value of $i=78^o$ best
fit the emission H$\alpha$ profiles. All fits were then rerun with
the inclination fixed at $78^o$. Thus the best-fit model for each
observed profile gives values for $\sigma_0$, $m$, and
$R_d$ at that date.

Figures~10 through 12 show the individual best-fits from 24~Aug 2004 through
1~Jan 2015. Overall the quality of the fits is good with the peak
separations, central depths, and overall widths of the H$\alpha$
profile generally reproduced. Close examination of the fits reveal
low-level asymmetries and core emission, that cannot be captured
by the models because the {\sc bedisk}/{\sc beray} modelling
assumes an axisymmetric disk and the synthetic H$\alpha$ profiles
are symmetric about line centre. Also shown in each panel of
Figures~10 through 12 is the model photospheric H$\alpha$ profile corresponding
to the star alone. 

Figure~13 summarizes all of the H$\alpha$ line profile modelling. Shown
is the model fit to the median profile of each identified state of EM~Cep (B, Be low,
Be middle, and Be high) given in Table~2. The disk parameters corresponding
to low, middle and high Be states are given in Table~5.
The obtained ranges of values of $\sigma_0$, $m$, and $R_d$ for the different states 
of the targets are within those typically found for the Be stars (Silaj et al.\
2010, Silaj et al.\ 2014).

The disk density parameters for each fit to an observed H$\alpha$
profile can be used to estimate the total mass in the H$\alpha$
disk as a function of observing date. The outer radius of the disk
used for these mass estimates is set to be $\min(R_d, 20 \,R_*)$
as the typical size of the H$\alpha$ formation region in large
disks is $\approx\,20$ stellar radii (Grundstrom $\&$ Gies 2006).
Because values for the index $m$ in Equation~\ref{eq:surfden}
that match the observed profiles
are large, $m\ge 2.5$, the resulting disk masses are not particularly
sensitive to the exact value of outer disk radius assumed.
To get some estimate of the uncertainties involved, disk masses
were estimated for the top 5 fitting H$\alpha$ line profiles for
each observed line (i.e.\ the 5 models with the lowest $\cal F$).
The symbol at each date is the mean disk mass with a $1\sigma$
variation shown.

\begin{figure}
\plotone{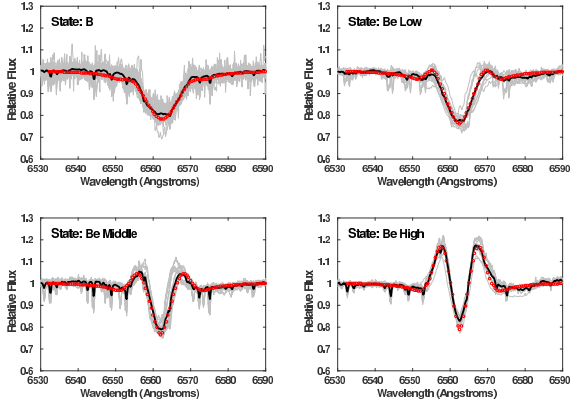}
\caption{Summary of H$\alpha$ fits according to disk state. In each panel,
the H$\alpha$ line profiles corresponding to the given state are shown
as light grey lines; the dark solid line is the median profile, and the red
circles are the best-fit model profile with disk parameters given in Table~3.
\label{fig:halpha_state}.}
\end{figure}

\begin{figure}
\begin{center}
\plotone{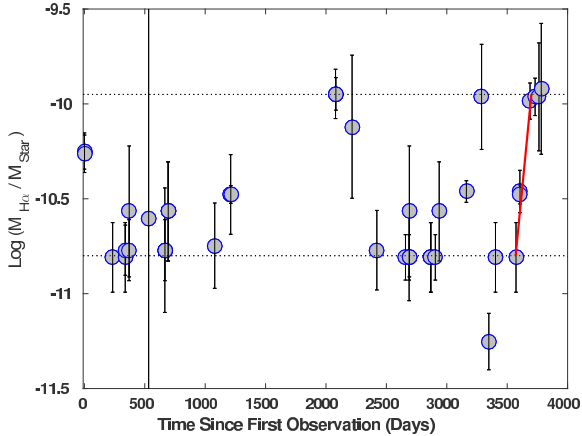}
\caption[]{Mass of the H$\alpha$ disk of EM~Cep as a function of
time from the best-fit density models. Shown for each date are the
mean and $1\sigma$ variation of the masses from the top 5 fitting
H$\alpha$ profiles. The horizontal dotted
lines give the approximate values for the high and low mass disk
states. The solid red line connects the 3 dates used to estimate the
mass loss rate during this period (see text).
\label{fig:halpha_disk_mass}}
\end{center}
\end{figure}

Figure~14 shows the H$\alpha$ disk mass of EM~Cep as a function of
time for the entire observational data set. It is seen to vary by
about an order of magnitude, between a low state of
$\log(M_D/M_*)<\approx -11.0$ and Be state of
$\log(M_D/M_*)\approx -10.0$. Only the exclusive low disk mass of
$\log (M_d/M_*) = -11.25$ for 24~Oct~2013 deviates considerably
from the rest values.

Figure~14 can also be used to estimate the mass-loss rate for EM~Cep.
Between the beginning of June 2014 and the end of September, 2014, the
mass of the disk increases by about $\approx\,10^{-9}\;\rm M_{\odot}$ giving a mass-loss
rate of approximately $\approx\,3\times 10^{-9}\;\rm M_{\odot}\,yr^{-1}$.
This value is roughly consistent with the range of values found observationally.
Based on the IR excess of Be stars, Waters et al.\ (1987)
estimated mass-loss rates (based on a simple, pole-on disk model) of between 
$\approx\,7\times 10^{-9}$ and
$\approx\,2\times 10^{-8}\;\rm M_{\odot}\,yr^{-1}$. Based on hydrodynamical modelling
of the optical light curve of the Be star 28~CMa, Carciofi et al.\ (2012) 
estimate a mass-loss rate of $3.8\times 10^{-8}\;\rm M_{\odot}\,yr^{-1}$.
Theoretically, Granada et al.\ (2013) suggest mass-loss rates in the range of
$10^{-11}$ to $10^{-7}\;\rm M_{\odot}\,yr^{-1}$ for stars losing mass during
episodes of critical rotation.

\begin{table}
\begin{center}
\caption[]{Disk parameters for the Be states of EM Cep. \label{t4}}
\begin{tabular}{lllll}
\hline\hline
State        & $\sigma_0$       & $m$ & $R_d/R_{\rm eq}$  & $M_d/M_*$ \\
             & ($\rm g\,cm^{-2}$) &     &  &  \\ \hline
Low Be       & $0.333$          & 3.0 & 12.0   & $3\times 10^{-11}$  \\
Middle Be    & $0.333$          & 2.5 & 25.0   & $5\times 10^{-11}$  \\
High Be      & $0.666$          & 2.5 & 25.0   & $1\times 10^{-10}$  \\ \hline
\end{tabular}
\end{center}
\end{table}

\section{Conclusions}

The main results of our study of EM~Cep in its B and Be state may
be summarized as follows:

\begin{enumerate}

\item The different levels of H$\alpha$ emission were formally
separated as initial, middle and high Be states. We detected that
the H$\alpha$ emission increased considerably in the framework
of several days at the highest Be state.

\item Eleven years of spectral observations, 2004-2015, reveal
that the target stay in the B state exceeds that in the Be state
and that the transition from B to Be to B state again lasted up to
7 months.

\item The spectra of EM~Cep within a single night do not change
considerably.

\item The H$\alpha$ profile of EM~Cep in the Be state is
doubly-peaked with a central absorption core and two emission
wings. The ratio of the emission peak flux and the central core
flux in the Be state nominally identify EM~Cep as a Be shell star.  The
separation between the emission peaks varied in the range
12--$15\;$\AA\ during the seasons. The H$\alpha$ profiles reveal
weak long-term ``V/R'' variability.

\item The spectral lines in our spectral range do not reveal any
apparent Doppler shifts in both the B and Be state.

\item The FWHM of the absorption H$\alpha$ line in the B state is
several times larger than that of the absorption core of the
H$\alpha$ line at Be state, while their depths are the same. The
total width of the whole H$\alpha$ profile at Be and B states are
almost the same.

\item The profiles of the absorption line He\,{\sc i}~6678 exhibit more
remarkable short-time variability than H$\alpha$. We report the
first detection of weak emission in the He\,{\sc i}~6678 line during the
highest Be state in May 2010.

\item We established the widening and red-shift of the He\,{\sc i}~6678
line in the phase range 0.754-0.874 during which the position of
the H$\alpha$ line remained constant.

\item We found similarity in widths and emission peak separations
of the H$_{\alpha}$ line of EM Cep at Be state and those of the
disk-like star FK Com and cataclysmic nova-like star UX UMa. This
may imply existence of gaseous disks with similar parameters for
different types of objects.

\end{enumerate}

Finally, the classical Be star model can satisfactorily
reproduce the H$\alpha$ line profiles of EM Cep with disk density parameters
and disk masses that are typical of the Be stars. In its high Be state, the
mass of EM Cep's H$\alpha$ disk is $\approx\,10^{-10}\,M_{*}$ and a disk
ejection episode observed in 2014 was estimated to imply a mass-loss rate
of $\approx\,3\times 10^{-9}\;\rm M_{\odot}\,yr^{-1}$.

\section*{Acknowledgements}

The research was supported partly by funds of project RD-08-81 of
Shumen University. TAAS wishes to acknowledge support from the
Canadian Natural Sciences and Engineering Research Council (NSERC).

{\it Facilities:} \facility{National Astronomical Observatory at Rozhen,
Bulgaria}

\appendix

%
%

\end{document}